\documentclass{article}
\usepackage{makeidx}  
\usepackage[utf8]{inputenc}
\usepackage[T1]{fontenc}
\usepackage{amssymb}
\usepackage{graphicx}
\usepackage{booktabs}
\usepackage{todonotes}
\usepackage{subfigure} 
\usepackage{multirow}
\usepackage{caption}
\usepackage{stmaryrd}
\usepackage{eurosym}
\usepackage{url}
\usepackage{natbib}
\usepackage{tablefootnote}
\usepackage[hang]{footmisc}
\usepackage{multirow}
\usepackage{natbib}

\begin{document}

\title{Collecting Examples for Instance-Spanning Constraints}

 \author{Stefanie Rinderle-Ma, Manuel Gall, Walid Fdhila, \\ J\"urgen Mangler, Conrad Indiono\\
University of Vienna, Faculty of Computer Science \\Research Group Workflow Systems and Technology\\
stefanie.rinderle-ma@univie.ac.at,\\ manuel.gall@univie.ac.at, \\walid.fdhila@univie.ac.at,\\ juergen.mangler@univie.ac.at, \\conrad.indiono@univie.ac.at}

\date{}
\maketitle              

\begin{abstract}
This report presents a meta analysis of various sources from literature, research projects, and experience with the goal of collecting examples for instance-spanning constraints to be implemented through Process-Aware Information Systems. 
\end{abstract}

\section{Introduction}
\label{sec:intro}
The goal of CRISP is the acquisition, formalization, implementation, monitoring, enforcement, and adaptation of instance-spanning constraints (ISC) in process-aware information systems (PAIS). ISC are a crucial building block for process-oriented applications, but have not been comprehensively investigated yet. An example requirement for ISC is the realization of bundling or unbundling of cargo over several logistic processes. 

In this report, ISC examples are collected based on a meta analysis of various sources. These sources include literature as well as various projects from different application areas. The ISC example collection provides the basis for a subsequent requirements analysis on ISC definition and implementation in PAIS. 

The report is structured as follows: In Section \ref{sec:collecting}, the methodology employed for ISC collection is outlined. Section \ref{sec:examples} presents the results of ISC collection. Section \ref{sec:summary} provides a summary and outlook. 

\section{Collecting ISC examples: Methodology}
\label{sec:collecting}
This section describes how sources and application areas were identified and contributed to this report.


\subsection{Related Work Analysis}
\label{sub:analyzerelwork}
We started from looking at the ``core'' related work which focuses on research in the area of instance-spanning, inter-instance, or cross-instance constraints, i.e., \cite{HealthCare_Heinlein2000,Formalism_iUPC_basic,DBLP:conf/emisa/Rinderle-MaM11,Security_Warner2006,OrderManagement_Pufahl2014,Leitner_ISC_2012,Many_Ly2015}. These approaches yield the starting point for further analysis in two ways, i.e., provide first examples and help to identify relevant application areas (see next step).  

\subsection{Application Areas Identification}
The identification of application areas was complemented by experiences from selected projects, i.e., the EU FP7 project ADVENTURE\footnote{\url{http://www.fp7-adventure.eu}, last access 2016-02-15} and the FFG project PROMISE\footnote{\url{http://cs.univie.ac.at/project/promise}, last access 2016-02-15}. Related application areas were identified based on core literature (cf. Section \ref{sub:analyzerelwork}) and the availability of information. The goal was to identify at least four different areas in order to obtain a sufficiently broad experience base. In the end, five areas were identified, i.e., health care (e.g., \cite{HealthCare_Heinlein2000,DBLP:conf/emisa/Rinderle-MaM11,Leitner_ISC_2012}), security (e.g., \cite{Security_Warner2006}), transport / logistics (e.g., \cite{OrderManagement_Pufahl2014}), manufacturing (e.g., \cite{Many_Ly2015}), and energy (cf. PROMISE project). 

\subsection{Document Collection for Application Areas}
\label{sub:collectareas}

For this step sources of real-world examples from different application areas had to be identified. Searching documents of application-oriented projects focusing on a specific area such as funded by the European Union (EU)\footnote{\url{http://ec.europa.eu/programmes/horizon2020/en}, last access 2016-02-15} seemed a feasible source, in particular, as all of the five identified areas health care, security, transport / logistics, manufacturing, and energy constitute pillars of the EU programme. Another source is the FFG\footnote{\url{www.ffg.at}, last access 2016-02-15} as Austrian funding schema for application-oriented research. In order to be comprehensive the FWF\footnote{\url{www.fwf.ac.at/en/}, last access 2016-02-15} as the Austrian funding scheme for basic research as well as the WWTF\footnote{\url{ww.wwtf.at}, last access 2016-02-15} as funding body for basic research projects with application prospects in Vienna were scanned for related projects as well. 
\begin{enumerate}
\item \emph{Identify relevant EU projects:} Relevant EU projects were identified based on the following methodology. At first, FP7 and H2020 project databases\footnote{\url{http://www.cordis.europa.eu/fp7/home_en.html}, last access 2016-2-15} were searched starting with \texttt{COOPERATION} projects. This selection of projects was refined based on the five application areas, i.e.,  \texttt{HEALTH},  \texttt{SECURITY},  \texttt{TRANSPORT}, \texttt{INTELLIGENT MANUFACTURING}, and  \texttt{ENERGY}. Note that \texttt{INT\-ELL\-I\-GENT MANUFACTURING} was not directly located under \texttt{CO\-OPERATION}, but under  \texttt{Other R\&D Programmes}. It was crucial to select ICT projects as cross-cutting concern in order to retrieve relevant results for PAIS. The search was also conducted in the opposite direction, i.e., starting with ICT projects and then selecting the five application areas. From these projects at the interface of ICT and the five application areas, were selected the ones that seemed to have a focus on processes by looking through the short project description.
\item \emph{Identify relevant FFG projects:} The focus was put on the ICT of the future (IKT der Zukunft) programme\footnote{\url{https://www.ffg.at/en/ictofthefuture}, last access 2016-02-15} as it reflects cooperative projects between research institutions and industry. Moreover, pillars of the programme relate to the identified application areas. 
Specifically, the database for {\tt Factories of the Future} projects\footnote{\url
{http://www.fabrikderzukunft.at/results.html/id5764}, last access 2016-02-15} was regarded a promising source. 
\item \emph{Identify relevant FWF projects} using the FWF project database\footnote{\url{https://pf.fwf.ac.at/en/research-in-practice/project-finder}, last access \\ 2016-02-15}. 
\item \emph{Identify relevant WWTF projects} using the WWTF project database\footnote{\url{http://www.wwtf.at/projects/research_projects/index.php?lang=EN}, \\ last access 2016-02-15} 
\end{enumerate}

\subsection{Analysis of industrial and experience papers from conferences in the business process domain}
Often conferences feature some industry or experience track where case studies are presented. We regard this as valuable source for ISC examples.
The conference selection is based on the process-orientation, i.e., business process management should be one of the relevant topics of the conference.
The international conferences on Business Process Management (BPM), Advanced Information Systems Engineering (CAISE), Service-Oriented Computing (ICSOC), Business Informatis (CBI), and Enterprise Computing (EDOC) were chosen. 

\subsection{Additional ISC examples from experience}
The motivation behind using ISC examples from our own experience, is that the authors have worked in different areas such as insurance, project management, and higher education. These examples from areas different from the five primarily selected ones can be seen as valuable extension reflecting the wide-spread use and importance of ISC. 






\section{Results of ISC collection}
\label{sec:examples}

In this section, the searches and results from the different sources described in Section \ref{sec:collecting} are presented. 

\subsection{Results from literature}
\label{sub:resultscore}
Table \ref{tab:iscexamlit} summarizes the ISC examples collected from the core literature as described in Section \ref{sub:analyzerelwork}. 
\begin{table}
	\centering
  \footnotesize
			\caption{ISC examples collected from literature with focus on instance-spanning constraints}
		\label{tab:iscexamlit}
		\begin{tabular}{p{1cm}p{1.7cm}p{0.5cm}p{9cm}}
    \toprule
    \textbf{Source} & \textbf{Area} &  \textbf{ID} &  \textbf{ISC example} \\ \midrule

\cite{HealthCare_Heinlein2000}  & Health care & (1) & 
 \texttt{\begin{tabular}{@{}p{9cm}} ``activities prepare 2, inform, call, and perform must be synchronized somehow as they access a ``limited resource,'' that is, the patient under consideration''
 \end{tabular}}\\ \hline
\cite{Formalism_iUPC_basic} & Health care & (2) & 
 \texttt{\begin{tabular}{@{}p{9cm}} ``Wait until centrifuge is filled.''
 \end{tabular}}\\ \hline
\cite{Printing_Pflug2013} & Health care & (3) &  
 \texttt{\begin{tabular}{@{}p{9cm}} ``Print similar jobs together.''
 \end{tabular}}\\ \hline
\cite{Many_Ly2015} & Manufacturing & (4) &   \texttt{\begin{tabular}{@{}p{9cm}} ``The carbon footprint of a supplier must not exceed a value of x. Depending on the number of suppliers modeled as resources, the constraint is instantiated multiple times. If suppliers can be added during runtime, the number of constraint instantiations will increase accordingly.''
 \end{tabular}}\\ \hline

&  & (5) &  \texttt{\begin{tabular}{@{}p{9cm}} ``There should not exist a group of n (say 2) users represented by G and m instances (say 4) such that if t1 is executed by a user in G, t2 is executed by the remaining users in G in more than m instances.'' 
 \end{tabular}}\\ \\

&  & (6) &  \texttt{\begin{tabular}{@{}p{9cm}} ``A user is not allowed to do t2 if the total loan amount per day exceeds \$1M.'' 
 \end{tabular}}\\ \\

\cite{Security_Warner2006} & Security & (7) &\texttt{\begin{tabular}{@{}p{9cm}} 
``There should not exist more than one instance of W such that the input parameters (say loan customer) is the same and the loan amount sums up to \$100K during a period of one month.'' \end{tabular}} \\  \\

&  & (8) &  \texttt{\begin{tabular}{@{}p{9cm}} ``There should not exist more than 4 instances of W such that a specific input parameter (say loan customer) is the same and the user executing t2 is also the same.'' 
 \end{tabular}}\\ \\

&  & (9) &  \texttt{\begin{tabular}{@{}p{9cm}} ``A user is not allowed to execute more than 100 tasks (of any workflow) in a day.'' 
 \end{tabular}}\\  \hline

\cite{OrderManagement_Pufahl2014} & Logistics & (10) &  \texttt{\begin{tabular}{@{}p{9cm}} ``For the synchronization of orders by the same customer and shipping address, ...''
 \end{tabular}}\\ \hline

      \bottomrule
		\end{tabular}
\end{table}

\subsection{Results from searching EU projects}
\label{sub:resultseuprojects}

Table \ref{tab:searchresults} summarizes the search results from European projects, i.e., those projects that focus on ICT and have one of the five application areas. The search was refined with criteria such as availability of information and service or process orientation.

\begin{table}
	\centering
  \footnotesize
		\caption{Search results on EU projects}
		
		\label{tab:searchresults}
		\begin{tabular}{p{2cm}p{4.0cm}p{5cm}}
    \toprule
      \textbf{Area} & \textbf{Search} & \textbf{Projects} \\ \midrule

	\emph{Health care} & refinement:\tablefootnote{\url{http://ec.europa.eu/research/health/index.cfm?pg=projects}} $+$ selection term \texttt{ICT}: FP6/FP7/H2020  & DECI: \url{http://deci-europe.eu/}; \newline MyCyFAPP: \url{http://www.mycyfapp.eu/en/}; \newline HEARTEN: \url{http://www.hearten.eu/}; \newline NoHoW: \url{http://nohow.eu/}  \\\hline
	
 		\emph{Security} & refinement: catalogue \cite{EuropeanUnion2014} & EDEN; CAPER; PREEMPTIVE; SECCRIT; EUROSKY; STAR-TRANS; IMCOSEC; ABC4EU, FASTPASS; CONTAIN; BRIDGE;  EMILI; PARIS; COMPOSITE; IDIRA (see description of all projects in \cite{EuropeanUnion2014}) \\\hline
 		
	\emph{Transport} & ICT-TRANSPORT with subject \texttt{ICT} and FP7-ICT with subject \texttt{TRANSPORT}; complement from mini cluster with ADVENTURE & E-FREIGHT: \tablefootnote{\url{http://www.efreightproject.eu/}}, EURIDICE: \tablefootnote{\url{http://www.euridice-project.eu/}}; GETSERVICE: \tablefootnote{\url{http://getservice-project.eu/}}\\\hline
	
	\emph{Manufacturing} & FP7-ICT with suject \texttt{Industrial Manufacture }& VISTRA: \tablefootnote{\url{http://www.vistra-project.eu/cms/htdocs/index.php?id=244}}; MSEE: \tablefootnote{\url{http://www.msee-ip.eu/project-overview}}; LinkedDesign: \tablefootnote{\url{http://www.linkeddesign.eu/}}; COMVANTAGE: \tablefootnote{\url{http://www.comvantage.eu/}}; ADVENTURE: \tablefootnote{\url{http://www.fp7-adventure.eu}}; VENIS: \tablefootnote{\url{http://www.venis-project.eu}}; FFD: \tablefootnote{\url{http://www.future-fashion-design.eu/TheProject.aspx}}; IMAGINE: \tablefootnote{\url{www.imagine-futurefactory.eu}}; EPES: \tablefootnote{\url{http://www.epes-project.eu/}}; GloNet: \tablefootnote{\url{www.glonet-fines.eu}}; PREMANUS: \tablefootnote{\url{www.premanus-project.eu}} \\\hline
	
	\emph{Energy} & ICT with subject \texttt{Energy saving} and ICT with subject \texttt{Energy transport and storage}& ISES: \url{http://ises.eu-project.info}\\ \midrule
	
	\multicolumn{3}{l}{availability of information, ICT aspect, service or process orientation,}\\
	\multicolumn{3}{l}{soft criterion: large scale}\\
	  
      \bottomrule
		\end{tabular}
\end{table}

Table \ref{tab:iscexameu} summarizes the ISC examples collected from the EU projects summarized in Table \ref{tab:searchresults}. Note that the ISC examples are numbered consecutively (as for all following tables containing ISC examples). 

Projects DECI, MyCyFAPP, NoHoW, EDEN were obviously just started in H2020, hence no public deliverables are available yet. For projects, CAPER, ABC4EU, IDIRA, PREMANUS, COMVANTAGE, LinkedDesign, FFD, ISES, VISTRA and CONTAIN, no relevant documents containing use case descriptions were available from the website. For STAR-TRANS and IMCOSEC the website was not available while for PREEMPTIVE, EPES, E-Freight and VENIS processes were available, but no explicit ISC were extractable. IMAGINE offered no deliverables in English. SECCRIT offers data protection requirements on a too coarse level for identifying ISC examples. As a consequence, it cannot be determined whether the requirement concerns instance-spanning aspects. For FASTPASS, no ISC were extractable. Finally, EUROSKY and COMPOSITE had a different scope than originally expected.

One interesting observation is that security projects yielding ISC examples are often more specifically residing in the area of crisis management, e.g., BRIDGE and EMILI.

\begin{table}
	\centering
  \scriptsize
			\caption{ISC examples collected from EU projects - Part 1}
		\label{tab:iscexameu}
		\begin{tabular}{p{1.5cm}p{1.5cm}p{0.5cm}p{8cm}}
    \toprule
    \textbf{Project} & \textbf{Area} &  \textbf{ID} & \textbf{ISC example\tablefootnote{For some projects we refer to specific deliverables in the form of Dx.x}} \\ \midrule    


\multirow{7}{*}{HEARTEN} & \multirow{7}{*}{Health care} & \multirow{3}{*}{(11)} & \texttt{``FR10: Alerts, reminders and motivation messages management'': ``Alerts and reminders are effectively managed and prioritized'' (cf. deliverable D3.1)} \\ \\
& & \multirow{3}{*}{(12)}& \texttt{``FR11: Patients follow up module'': ``At least one patient with regarded information should have been registered in the system.'' (cf. deliverable D3.1) }\\
\hline

BRIDGE & Security & (13) & \texttt{\begin{tabular}{@{}p{8cm}}``120 people in a train with one person doing triage at 30 seconds each, that's an hour of triage. While if you had a 5 second system, that brings it down to 5 minutes.'' (cf. D02.3) \end{tabular}}\\ \hline 
EMILI  & Security & (14) & \texttt{\begin{tabular}{@{}p{8cm}}``In a case where multiple alarms are detected in the same area and in a short time interval, the alarm is considered certain and fire detection is confirmed automatically.'' (cf.  \cite{ISSASIM-RDSS-camera-ready-Vuk-Valentina}) \end{tabular}} \\ \hline

\multirow{9}{*}{PARIS} & \multirow{9}{*}{Security} & \multirow{2}{*}{(15)} & \texttt{``A reasonable error rate for the system is 20\% of false recognitions'' (cf. D6.1)} \\ \\
& & \multirow{3}{*}{(16)}& \texttt{``The access to the APDB shall be able to store templates from at least 10 people.'' (cf. D6.1) }\\ \\
& & \multirow{3}{*}{(17)}& \texttt{``The system should implement adequate security measures to prevent or mitigate a denial of service attack'' (cf. D6.1)}\\
\hline

  &  & (18) & \texttt{\begin{tabular}{@{}p{8cm}}``For cargo distributed over several trucks, all cargo must arrive in the same destination. (cf. \cite{GetService1})'' (cf. D6.1)\end{tabular}} \\ \\ 
	
  &  & (19) & \texttt{\begin{tabular}{@{}p{8cm}}``For cargo distributed over several trucks, all cargo must arrive in the same destination, even the cargo is shifted to a partner transport organization.'' (cf. \cite{GetService2})\end{tabular}} \\

GET \newline SERVICE   & Transport &  &  \\

  &  & (20) & \texttt{\begin{tabular}{@{}p{8cm}}``If weather conditions change, a transport plane might land at another airport leading to rebundling of cargo.'' (cf. \cite{GetService1}) ) \end{tabular}} \\ \\ 
	
  &  & (21) & \texttt{\begin{tabular}{@{}p{8cm}}``Unloading part of the goods in order to continue the transport on the Danube. The unloaded goods can be sent by truck, train or another inland vessel.'' (cf.D1.1) ) \end{tabular}} \\ \hline

\multirow{11}{*}{EURIDICE} & \multirow{11}{*}{Transport} & \multirow{3}{*}{(22)} & \texttt{``Prioritization and dynamic handling of cargo by cargo-vehicle interaction to ensure high priority cargo item precedence over low priority items.'' (cf. \cite{EuridiceConsortium2009}) } \\ \\
& & \multirow{7}{*}{(23)}& \texttt{``Real time detection of exceptions in terms of missing boxes or delays that can trigger changes to the production orders and plans. The Intelligent Cargo is able to detect and inform the user about deviations during the transport process that can bring changes on the production order to which the cargo is assigned and on the future production plans.'' (cf. \cite{EuridiceConsortium2009})}\\
\hline

GloNet & Manufacturing & (24) & \texttt{\begin{tabular}{@{}p{8cm}}``As such besides the periodic maintenance of the plants, event-based maintenance also occurs when and if an abnormality is recorded by the monitoring system of the plant.'' (cf. D1.1)\end{tabular}} \\ \hline
MSEE & Manufacturing & (25) & \texttt{\begin{tabular}{@{}p{8cm}}``The vision for the TV sets is that the users can actively participate into content creation and can formulate the overall value and experience that each customer may enjoy.'' (cf. D1.1)\end{tabular}} \\ \hline

      \bottomrule
		\end{tabular}
\end{table}

\subsection{Results for searching FFG, FWF, and WWTF projects}
\label{sub:resultsffgprojects}

For FFG projects, the project database for \texttt{Factories of the Future} (cf. Section \ref{sub:collectareas}) was analyzed. Unfortunately, no ISC examples could be extracted. But a link to the FWF project database was found and therefore it was decided to scan through FWF projects as well. The search was done on \texttt{ALL} projects in area \texttt{102 Informatik} for the years from 2010 -- 2015 resulting in $427$ projects. Selecting those projects with a process focus resulted in $5$ projects. These projects either did not refer to constraints or had a different focus. 

The analysis also covered WWTF projects, where the focus was put on the ICT calls from 2008. Two projects with a focus on processes were selected. One project considered intra-instance constraints. The other project EDIMINE addresses Key Performance Indicators (KPIs) that are calculated over multiple instances and for which their performance has to be monitored \cite{Krathu2014}.

\begin{table}
	\centering
  \footnotesize
			\caption{ISC examples collected from WWTF projects}
		\label{tab:iscexamffg}
		\begin{tabular}{p{1.5cm}p{1.5cm}p{0.5cm}p{8cm}}
    \toprule
    \textbf{Project} & \textbf{Area} &   \textbf{ID} &  \textbf{ISC example} \\ \midrule

\multirow{24}{*}{EDIMINE} & \multirow{24}{*}{Logistics} &
\multirow{7}{*}{(26)} & \texttt{``Average ordered quantity per customer is used to measure customer satisfaction. The reason is that if a customer is not satisfied, ordered quantities are expected to be reduced as a consequence. We expect that the 3,750 per month. However, if it is less than half of the target value then it is considered critical.'' (cf. \cite{Krathu2014}) } \\ \\
& & \multirow{6}{*}{(27)}& \texttt{``Since INVOIC messages contain information of the line item monetary amounts, we can calculate this KPI by adding up all line item monetary amounts. We set the target value as 100,000. If it is less than 50,000, the KPI will be in critical state.'' (cf. \cite{Krathu2014}) }\\ \\
& & \multirow{4}{*}{(28)}& \texttt{``The optimal case is all deliveries are on-time or 100\% of on-time delivery. If the percentage of on-time delivery drops to 80\%, it is considered as critical.'' (cf. \cite{Krathu2014}) }\\ \\
& & \multirow{4}{*}{(29)}& \texttt{``We set two days as the critical threshold which means that if the majority of the actual deliveries arrive two days later or earlier than the requested date, this KPI will become critical.'' (cf. \cite{Krathu2014}) }\\

				\hline
      \bottomrule
		\end{tabular}
\end{table}

\subsection{Results from searching regulatory documents}
\label{sub:resultsregulatory}
The idea was to complement the search with an analysis of regulatory documents that have been identified as interesting sources during project analysis. 
One source for ISC examples from the energy / smart grid area discovered through the PROMISE project and the ``Austrian Use Cases for Smart Metering Advanced Meter Communication System (ACMS)'' (original in German: \"Osterreichs Use-Cases) \cite{oesterreichsenergie2015}. The document bundles all use cases for introduction of smart meters in Austria.

Another source for privacy regulations identified through the PARIS project\footnote{\url{http://paris-project.org/}, last access 2016-02-15} are the regulations analyzed in, for example, \cite{DBLP:conf/emisa/Rinderle-MaMM15}, i.e., The EDPS Video-Sur\-vei\-llance Guidelines\footnote{\url{https://secure.edps.europa.eu/EDPSWEB/webdav/shared/Documents/Supervision/Guidelines/10-03-17_Video-surveillance_Guidelines_EN.pdf}, last access 2016-02-15}, the OECD Privacy Guidelines\footnote{\url{http://www.oecd.org/sti/ieconomy/oecdguidelinesontheprotectionofprivacyand\%74ransborderflowsofpersonaldata.htm}, last access 2016-02-15}, the Guidelines for Public Video Surveillance\footnote{\url{www.constitutionproject.org/wp-content/uploads/2012/09/54.pdf}, last access 2016-02-15}, the Data protection and privacy ethical guidelines\footnote{\url{ec.europa.eu/research/participants/data/ref/fp7/89827/privacy_en.pdf}, \\last access 2016-02-15}, and the Operational Guidance on taking account of Fundamental Rights in Commission Impact Assessments\footnote{\url{ec.europa.eu/justice/fundamental-rights/files/operational-guidance_en.pdf}, last access 2016-02-15}.


Whereas the documents on privacy do not directly contain any ISC examples, the document on energy / smart grid processes includes several examples (more specifically service level agreements) that can be expressed and monitored by using ISC over the associated energy / smart grid processes. Tables \ref{tab:iscexamregdoc} and \ref{tab:iscexamregdoc2} show ISC examples taken from \cite{oesterreichsenergie2015}. Note that some of the ISCs have been aggregated, i.e., for the same ISC referring to different activities, one ISC containing all different activity options is considered. One example is ISC \texttt{For 100 (simultaneous) ad hoc readouts of end devices/activate/\allowbreak deactivate \allowbreak customer interface readouts/meter \allowbreak checks, 99\% $\leq$ 5 min is required.\allowbreak (min)\allowbreak (32)}. Here, all activities \texttt{ad hoc readouts of end devices}, \texttt{activate/\allowbreak deactivate customer \allowbreak interface readouts}, and \texttt{meter checks} must happen in at most 5 minutes for 99\% of all cases.

\begin{table}
	\centering
  \scriptsize
			\caption{ISC examples collected from regulatory documents: \cite{oesterreichsenergie2015}, domain energy -- Part 1}
		\label{tab:iscexamregdoc}
		\begin{tabular}{p{0.5cm}p{12cm}}
    \toprule
    \textbf{ID} & \textbf{ISC example} \\ \midrule
		(30) & \texttt{When starting the read-out of 00:00 values 99\% of all meters should be read out within 6 hours. (min)} \\
		(31) & \texttt{The \%-values of the SLA (line 5) have to be provided for 12 and 24 hours. (max) } \\
		(32) & \texttt{For 100 (simultaneous) ad hoc readouts of end devices/activate/deactivate customer interface readouts/meter checks, 99\% $\leq$ 5 min is required. (min) } \\
		(33) & \texttt{The ad hoc readout/activate/deactivate customer interface readout of single end devices requires 99\% $\leq$ 5 min for all cases. (min) } \\
		(34) & \texttt{In the ad hoc readout/"activate/deactivate customer interface" readout, the provider has to provide the \% values and times for 200, 500, and 1000 devices (simultaneously). (min) } \\
		(35) & \texttt{The  ad hoc read out/activate/deactivate customer interface readout of a single device has to be done in $\leq$ 2 min for 99\% of the cases. (max) } \\
		(36) & \texttt{Switching off/on a single device has to be realized in $\leq$ 5 min for 99\% of the cases. (min) } \\
		(37) & \texttt{For switching off/on, the provider has to provide the \% values and times for 200, 500, and 1000 devices (sequentially). (min) } \\
		(38) & \texttt{Switching off/on a single device has to be realized in $\leq$ 2 min for 99\% of the cases. (max) } \\
		(39) & \texttt{Order ``set limit for meter performance''/``set integration time for restricting meter performance'' has to be realized in $\leq$ 5 min in 99\% of the cases. (min) } \\
		(40) & \texttt{Order ``set limit for meter performance''/``set integration time for restricting meter performance'' for a single device has to be realized in $\leq$ 5 min in 99\% of the cases. (min) } \\
		(41) & \texttt{Order ``set limit for meter performance''/``set integration time for restricting meter performance''/``load control''/``activate or deactivate maintenance interface''/``deactivate communication interface at device''/``activate, deactivate prepayment'' has to be realized in $\leq$ 2 min in 99\% of the cases. (max) } \\
		(42) & \texttt{For 100 (simultaneous) orders ``control display..''/``display for customer change''/``control interval in customer interface''/``remote log..''/``parametrize rate..''/``put readout plan..''/``remote parametrization''/``set free text..''/``load control''/``activate, deactivate prepayment''/``load prepayment'', 99\% $\leq$ 5 min is required. (min) } \\
		(43) & \texttt{Order ``control display''/``display for customer change''/``control interval in customer interface''/``remote log..''/``parametrize rate..''/``put readout plan..''/``remote parametrization''/``set free text..''/``load control''/``activate, deactivate prepayment''/``load prepayment'' requires 99\% $\leq$ 5 min for all cases. (min) } \\
		(44) & \texttt{Order `control display''/``display for customer change''/``control interval in customer interface''/``remote log..''/``parametrize rate..''/``put readout plan..''/``remote parametrization''/``set free text..'', the provider has to provide the \% values and times for 200, 500, and 1000 devices (simultaneously). (min) } \\
		(45) & \texttt{For 100 (simultaneous) orders `control display''/``display for customer change''/``control interval in customer interface''/``remote log..''/``parametrize rate..''/``put readout plan..''/``remote parametrization''/``set free text..''/``load prepayment'', 99\% $\leq$ 2 min is required. (max) } \\
		(46) & \texttt{99\% of all devices/gateways have to show the same time and date (variance $<$ 2 sec) for each planned and ad hoc read out of the central system. (min) } \\
		(47) & \texttt{99\% of the devices/meters successfully receive their firmware in full mode within 20 days. (min) } \\
		(48) & \texttt{99,8\% of the gateways successfully receive their firmware within 24 hours (min). } \\
		(49) & \texttt{The provider has to provide the vales of a pure firmware update in full operation for 99\% of the devices/meters.(max) } \\
		(50) & \texttt{99,98\% of the gateways successfully receive their firmware within 24 hours (max). } \\
		(51) & \texttt{Order ``load activation time of a firmware upgrade''/``cancel firmware for devices''/``set remote load unit table''/``set remote meter switch table'' has to be done $\leq$ 24 hours in 99\% of the cases. (min) } \\
				\hline
      \bottomrule
		\end{tabular}
\end{table}

\begin{table}
	\centering
  \scriptsize
			\caption{ISC examples collected from regulatory documents: \cite{oesterreichsenergie2015}, domain energy -- Part 2}
		\label{tab:iscexamregdoc2}
		\begin{tabular}{p{0.5cm}p{12cm}}
    \toprule
    \textbf{ID} & \textbf{ISC example} \\ \midrule
		(52) & \texttt{The provider has to provide the \% values and times for 1000, 10000, 50000, 100000, and 1000000 simultaneous orders ``load activation time of firmware updgrade'' for a switch on. (min) }\\
		(53) & \texttt{Order ``load activation time of a firmware upgrade''/``cancel firmware for devices''/``set remote load unit table''/``set remote meter switch table'' has to be done $\leq$ 60 min in 99.9\% of the cases. (max) }\\
		(54) & \texttt{The provider has to provide the \% values and times for 1000, 10000, 50000, 100000, and 1000000 simultaneous orders ``cancel firmware for devices''. (min) }\\
		(55) & \texttt{The provider has to provide the \% values and times for 200., 500, and 1000 (simultaneously) sent ALARM or EVENT messages of meters. (min) }\\
		(56) & \texttt{Sending EVENTs of single meters is to be done $\leq$ 1 hour for 99\% of the cases. (max) }\\
		(57) & \texttt{For 100 (simultaneously) sent ALARM/EVENT messages  99\% $\leq$ 5 min is required. (min) }\\
		(58) & \texttt{Sending an ALARM/EVENT message by a single end device/gateway requires 99\% $\leq$ 5 min for all cases. (min) }\\
		(59) & \texttt{The provider has to provide the \% values and times for 200, 500, 1000, 10000, 50000, and 100000 (simultaneoulsy) sent ALARM/EVENT messages (min) }\\
		(60) & \texttt{Sending an ALARM message by a single end device/gateway requires 99\% $\leq$ 2 min for all cases. (max) }\\
		(61) & \texttt{Controlling load control unit has to be done $\leq$ 5 min for 99\% of the cases. (min) }\\
		(62) & \texttt{The provider has to provide the \% values and times for 100, 200, 500, and 1000 load control units. (min) }\\
		(63) & \texttt{Controlling load control unit has to be done $\leq$ 2 min for 99\% of the cases. (min) }\\
		(64) & \texttt{The provider has to provide the \% values and times for 200, 500, 1000, 10000. and 100000 for (simultaneous) orders ``load control''/``set remote load unit table''. (min) }\\
		(65) & \texttt{The provider has to provide the \% values and times for 1000, 10000, 50000, and 100000 simultaneous orders ``set remote meter switch table''. (min) }\\
		(66) & \texttt{Managing bottlenecks for all load units has ot be done within 30 seconds in 99\% of the cases. (min) }\\
		(67) & \texttt{For 200, 500, and 1000 (simultaneous) orders "activate or deactivate maintenance interface"/"deactivate communication interface at device"/"activate, deactivate prepayment", the provider has to provide the \% values and times. (min) }\\
		(68) & \texttt{The (first) registration/re-registrations of a device has to be done  $\leq$  6 hours for 99\% of the cases. (min) }\\
		(69) & \texttt{The provider has to provide the \% values and times for 100, 200, 500, and 1000 meters (sequentially) for (first) registration/re-registration of a device. (min) }\\
		(70) & \texttt{The (first) registration of a device has to be done  $\leq$  60 min in 99\% of the cases. (max) }\\
		(71) & \texttt{The re-registration of a device has to be done  $\leq$  15 min in 99\% of the cases. (max) }\\
		(72) & \texttt{The first registration of a gateway has to be done  $\leq$  30 minutes for 99\% of the cases. (min) }\\
		(73) & \texttt{The first registration of a gateway has to be done  $\leq$  5 minutes for 99\% of the cases. (min) }\\
		(74) & \texttt{The provider has to provide the \% values and times for 200, 500, and 1000 gateways/cryptographic changes/rights transmissions/changes at devices and gateways/orders to devices or gateways simultaneously. (min/max) }\\
		(75) & \texttt{For one gateway the provider has to provide the \% values and times  $\leq$  5 min in 99\% of the cases. (max) }\\
		(76) & \texttt{For 100 (simultaneous) devices for which cryptographic parameters were changed/for which rights were transmitted/ device and gateway changes/orders to devices or gateways, 99\% $\leq$  5 min is required. (max) }\\
		(77) & \texttt{For changing cryptographic parameters/transmitting rights/ changing single devices or gateways, 99\%  $\leq$ 5 min is required. (max) }\\
				\hline
      \bottomrule
		\end{tabular}
\end{table}

\subsection{Results from searching papers from industry and experience conference sessions}
\label{sub:resultsindustrypapers}

The Int'l Conference on Business Process Management (BPM) was scanned for industrial papers since 2000 resulting in $4$ industrial papers describing in total $13$ ISC examples (cf. Table \ref{tab:iscexamindusconf}).  
For the other conferences, only few industrial papers were found or they did not have a specific process focus. Sometimes industrial papers were more of tool descriptions than presenting real-world studies. 

\begin{table}
	\centering
  \scriptsize
			\caption{ISC examples collected from industry and experience conference sessions}
		\label{tab:iscexamindusconf}
		\begin{tabular}{p{1.5cm}p{1.5cm}p{0.5cm}p{8cm}}
    \toprule
    \textbf{Paper} & \textbf{Area} &  \textbf{ID} & \textbf{ISC example} \\ \midrule    
\cite{BPM2015_Paper3} & Insurance &  (78) & \texttt{\begin{tabular}{@{}p{8cm}} R0: ``acknowledgement\_of\_application.started occurs at least 1x''
\end{tabular}} \\ \hline

\multirow{27}{*}{\cite{BPM2014_Mrasek_etal}} & \multirow{27}{*}{Automotive} &
\multirow{2}{*}{(79)} & \texttt{``P3.1 Maximal UDS Connections: The number of connections to UDS should not exceed 10''} \\ \\
& & \multirow{2}{*}{(80)}& \texttt{``P3.2 Maximal KWP-2000 Connections The number of connections to KWP2000 should not exceed 10.'' }\\ \\
& & \multirow{2}{*}{(81)}& \texttt{``P3.3 Maximal Connections The number of connections UDS and KWP2000 should not exceed 14.'' }\\ \\
& & \multirow{2}{*}{(82)}& \texttt{``check if all its Electronic Control Units (ECU) are integrated correctly and to put the ECUs into service.'' }\\ \\
& & \multirow{2}{*}{(83)}& \texttt{``The number of false positives,  and the number of false negatives in the processes, should be small.'' }\\ \\
& & \multirow{4}{*}{(84)}& \texttt{``Some ECUs require specific resources at the process place for their testing. When a task requires a resource not available at the current process place the process is blocked.'' }\\ \\
& & \multirow{5}{*}{(85)}& \texttt{``Each transport protocol (UDS, KWP2000) can handle a certain number of open connections, in our environment 10 at the same time. In total, 14 connections altogether can be open at the same time. To avoid blocking of a process, the process must not open more connections.'' }\\ \\
& & \multirow{1}{*}{(86)}& \texttt{``Only one task at the same time can access/test each ECU C'' }\\ \\
& & \multirow{1}{*}{(87)}& \texttt{``Some ECU C must never be tested in parallel with an ECU C2'' }\\ \hline

 &  & (88) &
\texttt{\begin{tabular}{@{}p{8cm}} ``if (incident priority isLow) and if(not novice isAvailable) and if (expert isAvailable) $\longrightarrow $ wait in queue)''\end{tabular}}\\ \\
\cite{BPM2014_Sindghatta} & IT Incident Management & (89) &
\texttt{\begin{tabular}{@{}p{8cm}} ``if (incident priority isLow) and if(not novice isAvailable) and if (not expert isAvailable) $\longrightarrow $ wait in queue'' \end{tabular}}\\
 &  & (90) &
\texttt{\begin{tabular}{@{}p{8cm}} ``if (incident priority isHigh) and if(not expert isAvailable) and if (not novice isAvailable) $\longrightarrow $ wait in queue'' \end{tabular}} \\ \hline

\cite{BPM2012_Estruch} & Manufacturing & (91) &\texttt{\begin{tabular}{@{}p{8cm}}``Detect production line failure when consecutive bad finished products are found.''
\end{tabular}} \\ \hline

 &  & (92) &\texttt{\begin{tabular}{@{}p{8cm}}``At the end of iteration, award each contributor who scored better than the average score of his neighbors in that iteration.''\end{tabular}} \\
\cite{BPM2012_Skecic} & Social Computing & \\
 &  & (93) &\texttt{\begin{tabular}{@{}p{8cm}}``Assign the person with most check-ins at a place `Mayor' badge.''\end{tabular}} \\ \hline

\cite{InformationSecurity2012} & Security & (94) & \texttt{\begin{tabular}{@{}p{8cm}}``Sabre [...] is a system that analyzes the browser extensions by monitoring in-browser information-flow. It produces an alert when an extension is found to access some sensitive information in an unsafe way.''\end{tabular}} \\ \hline
      \bottomrule
		\end{tabular}
\end{table}

\subsection{ISC examples from experience}
\label{sub:resultsexperience}

Table \ref{tab:iscexamexperience} summarizes ISC examples collected from own working or project experience. Reflecting this diverse experience, the ISC examples also stem from other areas than the originally selected ones. However, most of the examples can be related to areas such as health care, security, manufacturing, transport / logistics, and energy / smart grids. 

\begin{table}
	\centering
  \scriptsize
			\caption{ISC examples from experience}
		\label{tab:iscexamexperience}
		\begin{tabular}{p{1.5cm}p{0.5cm}p{11cm}}
    \toprule
     \textbf{Area} &  \textbf{ID} &  \textbf{ISC example} \\ \midrule

\multirow{9}{*}{Law} & 
\multirow{2}{*}{(95)} & \texttt{Due to agreed payment all instances of the file will be closed immediately (within 24 hours). } \\ \\
& \multirow{2}{*}{(96)}& \texttt{Due to monthly payment all instances of all files from this customer will be set on hold immediately (within 24 hours).  }\\ \\
& \multirow{1}{*}{(97)}& \texttt{After a ZMR request another request is not allowed for 1 month (per customer). }\\ \\
& \multirow{1}{*}{(98)}& \texttt{An employee has to finish his critical tasks each day within his regular work hours. }\\ \hline

Collection Agency & (99) &\texttt{\begin{tabular}{@{}p{11cm}} An employee is allowed to give a customer a 20 percent discount on his file. The monthly average discount shall not exceed 10 percent. \end{tabular}} \\\hline
Teaching & (100) &\texttt{\begin{tabular}{@{}p{11cm}} When one student (instance) has not understood a concept, the concept is teached to every student in class again.\end{tabular}} \\\hline

\multirow{10}{*}{Manufacturing} & 
\multirow{2}{*}{(101)} & \texttt{During production process a blade got lost now all intances have to be searched until the blade is found. } \\ \\
& \multirow{2}{*}{(102)}& \texttt{There was a problem with a product and now all instances are recalled / checked. (e.g., supermarket, or in house quality testing) }\\ \\
& \multirow{4}{*}{(103)}& \texttt{A manufacturer produces different type of products based on customer orders. During manufacturing, the production of some products should be separated from the others (toxic products and food products). The storage task takes into consideration this constraints. }\\ \hline

\multirow{8}{*}{Logistics} & 
\multirow{2}{*}{(104)} & \texttt{If an instance of a transport process gets delayed e.g. border control, you may change other instances of this process to an other operator (ship, train).} \\ \\
& \multirow{2}{*}{(105)}& \texttt{When a Incident happens all instances of all processes routing this way have to be changed. }\\ \\
& \multirow{2}{*}{(106)}& \texttt{In a manufacturing process, the shipment of different goods can be optimized by for example batching things to be transported to the same place. }\\ \hline

\multirow{10}{*}{Health care} & 
\multirow{3}{*}{(107)} & \texttt{A patient is involved simultaneously in two different treatments, i.e., dermatology and diabetes. A list of drugs should not be taken during the dermatology treatment. The diabetes treatment should consider this list. } \\ \\
& \multirow{3}{*}{(108)}& \texttt{A patient is involved simultaneously in two different treatments, i.e., dermatology and diabetes. If certain treatments are already done by the diabetes analysis and should not be done again for the dermatology tests (skin infections). }\\ \\
& \multirow{2}{*}{(109)}& \texttt{In hospital, depending on the gravity of the case, the order of checking or treatment is defined. }\\ \hline

\multirow{5}{*}{Automotive} & 
\multirow{2}{*}{(110)} & \texttt{Several diagnosis for a car are running at the same time. If the problem is identified in one of the diagnosis, the others are cancelled.  } \\ \\
& \multirow{2}{*}{(111)}& \texttt{Several diagnosis for a car are running at the same time. Similar tasks in different checks should be only executed once. }\\ \hline

& (112) & \texttt{\begin{tabular}{@{}p{11cm}}
In a Doctor cabinet, several patients are making the same checks. A patient that have a contaminating illness (Ebola) can have effects on the other patients processes; e.g. supplementary checks if they are contaminated. \end{tabular}} \\

Crisis \newline management & & \\

& (113) & \texttt{\begin{tabular}{@{}p{11cm}}
Several planes are in the landing process. A plane that had a problem during the landing can affect the landing of the other planes. \end{tabular}} \\\hline

 Safety, \newline Security & (114) & \texttt{\begin{tabular}{@{}p{11cm}} Several people are accused for a same murder. For the interviewing process, these people should be separated and are not allowed to talk to each other. \end{tabular}} \\\hline

      \bottomrule
		\end{tabular}
\end{table}

\newpage
\section{Conclusions and Outlook}
\label{sec:summary}

\begin{figure}
\centering
\subfigure[\#ISC along areas]{\includegraphics[width=0.49\linewidth]{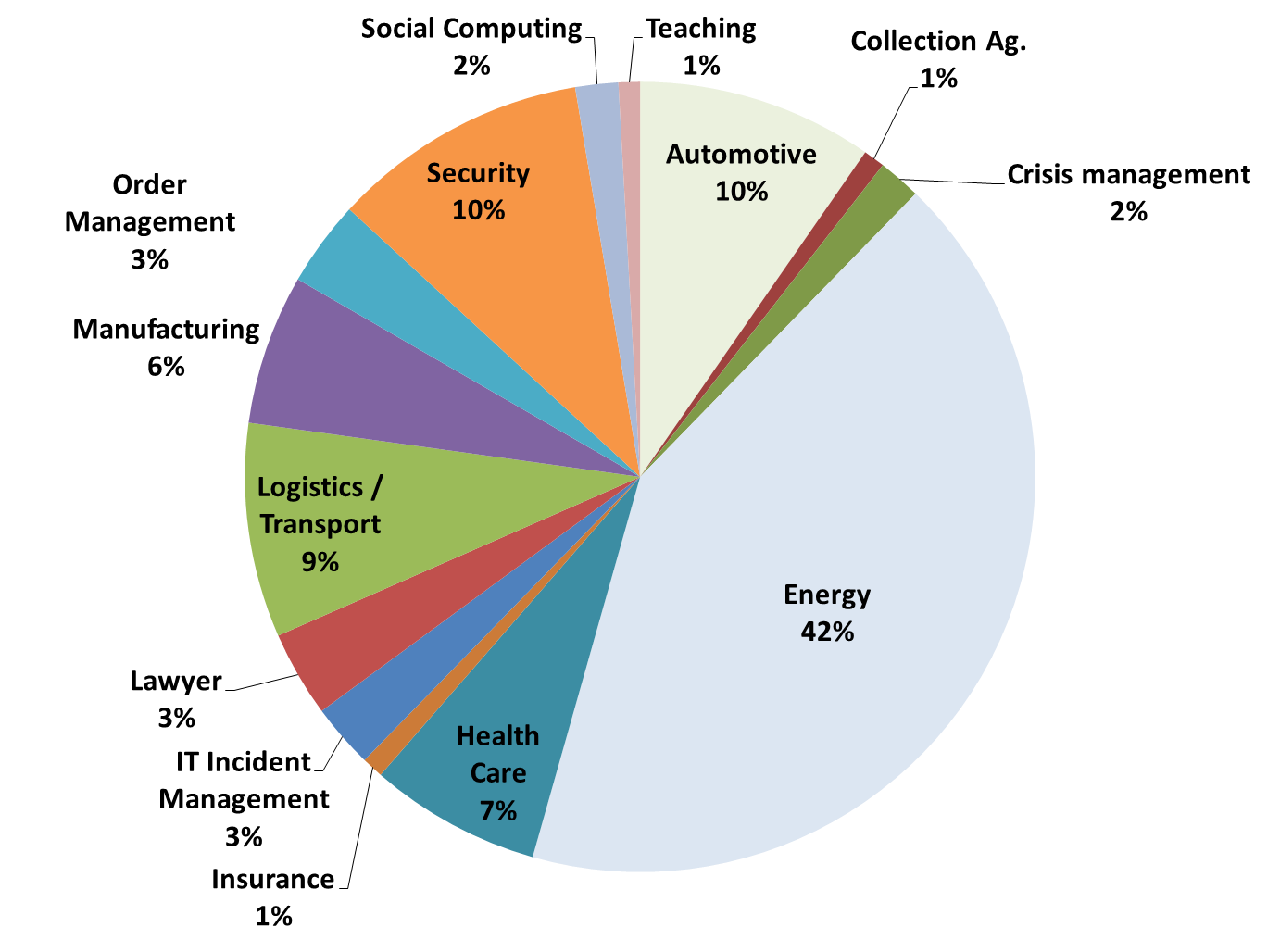}}
\subfigure[\#ISC along sources]{\includegraphics[width=0.49\linewidth]{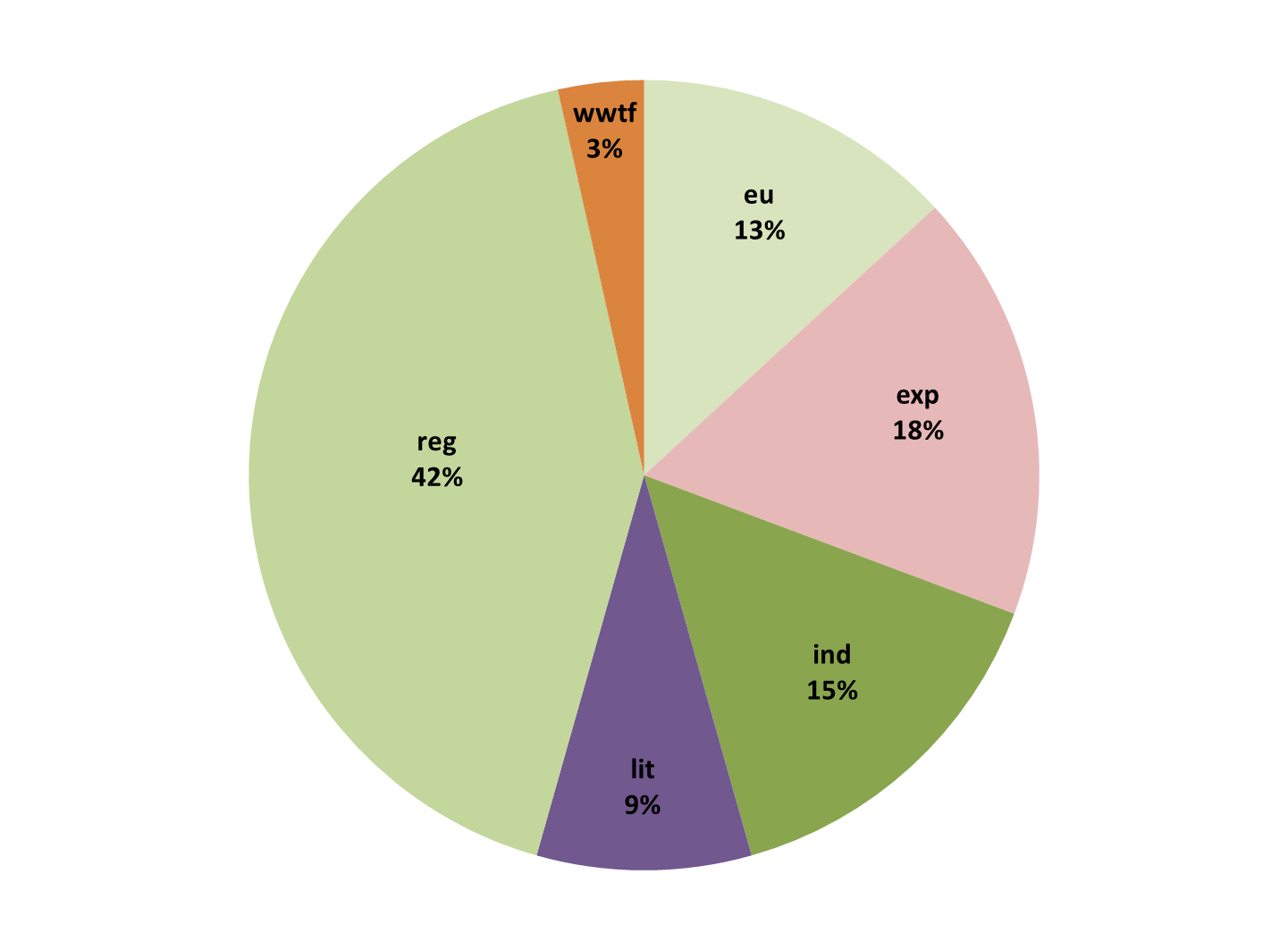}}
\caption{Distribution of ISC examples along areas and along sources}
\label{fig:distriscarea}
\end{figure}

In summary, $114$ ISC examples were collected during this meta analysis. Figure \ref{fig:distriscarea}(left) illustrates the distribution of ISC along the different application areas. It can be seen that all five originally selected areas seemingly demand for the support of ISC. The energy domain seems to be the most effective source for ISC examples, followed by security, logistics / transport, health care and manufacturing. If manufacturing is combined with automotive, the second highest number of examples stem from this combined category. Overall, the result is not really surprising as the search was tailored towards these five areas. An interesting category is automotive as it was not explicitly selected and searched for. The automotive domain seems to play an important role contributed by industry papers (ind) and experience (exp). 

On the right side of Figure \ref{fig:distriscarea}, the distribution of ISC examples along the analyzed sources is shown. The majority of ISC examples stem from regulatory documents (reg), specifically from a single document. The low number of analyzed regulatory documents (one for energy and five for security) is indeed a limitation because there is an abundance of regulatory documents for all kinds of areas in general. Hence, in future work, regulatory documents must be carefully selected and analyzed in order to come up with more ISC examples. 

We could think about grouping the ISC examples into one group literature (lit and ind) and the rest that accounts for the application of ISC, i.e., all projects, all regulatory documents, and all experiences. Then the distribution is 24\% examples from literature and 76\% ISC examples from application. It can be expected that the analysis of further regulatory documents might even increase the percentage of ISC examples from application. This might indicate that the demand from application towards the support of ISC is big. At the same time it is to be investigated whether existing approaches reflected by literature can respond to this demand. This is of course not decidable on basis of this meta analysis as only examples were considered and no approaches, but at least some recommendation to increase research on ISC can be made. 

\section{Acknowledgements}
\emph{This work has been funded by the Vienna Science and Technology Fund (WWTF) through project ICT15-072}.

\bibliography{CRISP}

\end{document}